# Plasticity of the Al₂Cu θ-phase from ambient temperature to 300°C

D. Andre*, J.S.K.-L. Gibson, P.-L. Sun, N. Lohrey, S. Sandlöbes-Haut, S. Korte-Kerzel

Institute of Physical Metallurgy and Materials Physics, RWTH Aachen University, Germany

*andre@imm.rwth-aachen.de (corresponding author)

## Abstract

The investigation of the deformation behaviour of intermetallic phases is mostly limited to high temperatures due to their low ductility at ambient temperature. Therefore, within this study, nanoindentation experiments on the Al₂Cu θ-phase were performed from ambient temperature up to 300°C in conjunction with TEM investigations of the deformed material. It was found that the Al₂Cu θ-phase starts to soften from temperatures above 150°C. The low temperature deformation behaviour was dominated by serrated yielding and transitioned towards a smooth deformation behaviour above 150°C. No slip traces were observed in the vicinity of the indents in both temperature regimes. The predominant dislocations observed after ambient temperature deformation and elevated temperature deformation are assumed to belong to the same operating slip systems. Therefore, the underlying deformation mechanism for both temperatures appears to be the same and thermally activated.

Keywords: Al₂Cu-θ phase, dislocations, temperature, nanoindentation, TEM

## 1. Introduction

Intermetallic phases often combine several advantageous properties such as a high strength and good corrosion resistance and are therefore of high interest for advanced applications. The intermetallic Al₂Cu θ-phase, including its metastable variants during precipitation, is mostly used as a strengthening phase embedded in an Al-matrix. Even though it is one of the most intensively studied intermetallic phases in the literature [1-6], its deformation behaviour is not completely understood yet. One of the main reasons for this is its limited ductility at ambient temperature.

So far, conventional macroscopic mechanical testing of the Al₂Cu θ-phase was performed by Chanda et al. [1] and Dey et al. [2]. Dey et al. [2] reported a brittle-to-ductile transformation temperature (BDTT) of 450°C in tension and 400°C in compression tests and assigned the plastic deformation to the Peierls-Nabarro mechanism. Chanda et al. [1] found a similar BDTT of 375°C by means of compression testing. Further, the deformation behaviour of the Al₂Cu θ-phase was investigated for an Al-Al₂Cu composite material. A study by Ignat et al. [4] revealed restricted plastic deformation up to 300°C and dislocation multiplication and glide at 350°C in the intermetallic phase using TEM investigations. They further reported cross slip at these elevated temperatures.

In order to overcome the difficulty to gain mechanical data at lower temperatures, conventional hardness testing or instrumented nanoindentation testing can be used, as the confining pressure of the stress state under an indent is known to suppress cracking [7]. Statistical indentation phase mapping allows not only the quantification of the hardness and elastic modulus of a particular crystallographic orientation but produces data over a wide range of crystallographic orientations and therefore enables an estimation of the degree of anisotropy of the material. This method was applied by Xiao et al. [8] on the Al₂Cu θ-phase at ambient temperature and revealed that the hardness of the phase mostly lies in the range between 5 GPa and 7 GPa and that the modulus lies mostly between 110 GPa and 130 GPa. Thus, the $Al_2Cu$ θ-phase has slightly anisotropic mechanical properties. Furthermore, nanoindentation data at room temperature as well as at 200°C and 350°C of the $Al_2Cu$ θ-phase with small additions of Ni were reported by Chen et al. [9]. The reported average hardness values amounted to 5.77 GPa at room temperature, 5.33 GPa at 200°C and 2.53 GPa at 350°C, whereas the reduced elastic modulus was reported to amount to 109.7 GPa at ambient temperature and 101.2 GPa at 200°C. Results obtained by the pulse-echo-overlap technique by Eshelman et al. [5] revealed an Young´s modulus value of 109.9 GPa at ambient temperature, whereas recent density functional theory (DFT) calculations performed by Zhang et al. [10] revealed Young´s modulus values of 120.1 GPa, corresponding to the Voigt-Reuss-Hill average.

Further studies on the activated plasticity mechanisms of the pure intermetallic phase are mostly conducted at elevated temperatures, whereas studies on the lamellar $Al$-$Al_2Cu$ eutectic allow to plastically deform the compound at ambient temperature. These studies revealed dislocations in the $Al_2Cu$ θ-phase on {110}, {200}, {112}, {011}, {001}, {010}, {111} and {211} planes [3, 4, 11-14] as well as localized shear on {011} planes and faulted structures on {211} planes [12]. A recent publication by the present authors [15] further took advantage of nanomechanical methods to access the plasticity of the $Al_2Cu$ θ-phase at ambient temperature. Micropillar compression allowed measurements of the critical resolved shear stresses of several slip systems and a new prediction of the order of easy to activate slip systems in terms of their critical resolved shear stress, which proved consistent with the observed slip systems {211}½<$\bar{1}$11>, {110}½<$\bar{1}$11>, {022}½<1$\bar{1}$1> and {022}<100>.

However, although these studies lead towards a better understanding of the mechanical properties and corresponding deformation mechanisms over a large range of temperature, much remains unknown. Therefore, in this work, we investigated the mechanisms governing plasticity over a wide range of temperatures of the $Al_2Cu$ θ-phase using nanoindentation at small temperature intervals in the range from room temperature to 300°C and transmission electron microscopy (TEM) to study the resulting microstructures after room temperature and elevated temperature indentation.

## 2. Experimental procedure

### 2.1. Room to high temperature nanoindentation experiments

The $Al_2Cu$-θ as-cast specimen was ground and polished in order to perform nanoindentation experiments at different temperatures, namely, at ambient temperature, 50°C, 100°C, 150°C, 200°C, 250°C and 300°C using a MicroMaterials NanoTest P-3 system, modified with a custom-built vacuum chamber ($\sim 10^{-4}$ mbar, see reference [16]) using a Berkovich tip made out of cBN (cubic boron nitride), supplied by Synton MDP, Switzerland. The tests were performed until a maximum load of 100 mN was reached at a loading and unloading rate of 10 mN/s. In order to reduce the effect of creep on the results, a holding period of 5 s at the maximum load was applied. For the experiments, the indenter tip and the sample were heated separately and an additional thermocouple was attached close to the sample surface in order to track the sample temperature [16, 17]. Before performing the actual experiments at high temperatures, indents to match the temperature to that measured on the sample were performed. All results therefore exhibit a post-indentation drift rate of less than 0.2 nm/s. The subsequently conducted data analysis is based on the Oliver and Pharr method, and the tip area function was calibrated using the same method on fused silica at room temperature prior to the main tests [18].

### 2.2. TEM investigations

To further investigate the underlying deformation substructure in the plastic zone underneath the indents, TEM analyses on indents made at room temperature (RT) and at elevated temperature (300°C) were performed. Therefore, TEM membranes with a final thickness of approximately 100-200 nm were prepared using a focussed ion beam milling (FIB, Helios Nanolab 600i, FEI) and a final polishing step at a low accelerating voltage of 5 kV. The lamellae were analysed using a JEM-F200 (JEOL Ltd.) with an acceleration voltage of 200 kV. However, since the indentation experiments were performed from ambient temperature up to elevated temperatures, the indents performed at room temperature were heated after indentation. In order ensure that there was no effect of heating on the microstructure under the indent, a further ambient temperature indent was performed and analysed in the TEM.

## 3. Results
### 3.1. Nanoindentation experiments

The deformation behaviour of the intermetallic Al$_2$Cu θ-phase was investigated by means of nanoindentation from ambient temperature up to 300°C. The resulting, thermal drift-corrected, load-displacement curves for all temperatures are depicted in Figure 1. At a temperature of 200°C the material begins to soften as evident from the lower slope of the loading curve and increased indentation depths at maximum load and during the holding period. This is also revealed in terms of hardness (Figure 2). Therefore, for the following analyses, temperatures up to 150°C are referred to as the low temperature regime and temperatures starting from 200°C are signified as the high temperature regime.

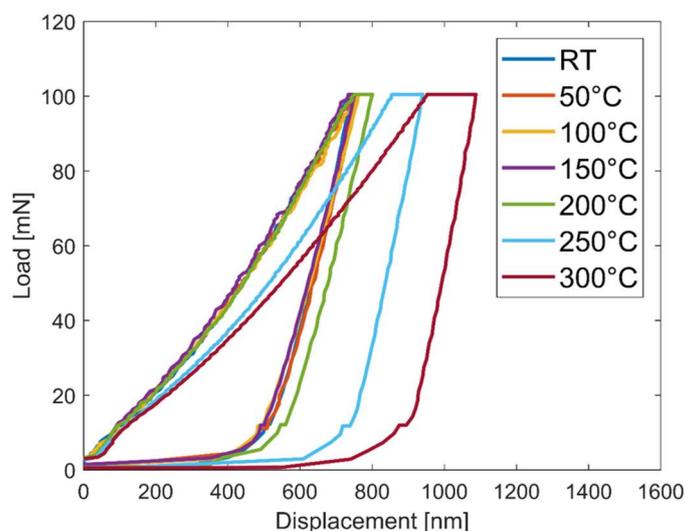

Figure 1: Nanoindentation load-displacement curves of the Al$_2$Cu θ-phase at different temperatures with a loading and unloading rate of 10 mN/s until a maximum load of 100 mN with a holding period of 5 s at maximum load. All curves are corrected for thermal drift.

Furthermore, the upper segment of the unloading curve was slightly bent towards higher displacements at high temperatures, presumably due to a small amount of residual creep deformation. The fit to the unloading curve, which was performed according to the Oliver and Pharr analysis [19], was therefore varied between 100-20% and 50-20% of maximal load to reveal its influence on the hardness. However, only slight deviations in hardness occurred (Figure 2) with the average hardness of the 50-20% fit being lower than the average hardness of the 100-20% fit.

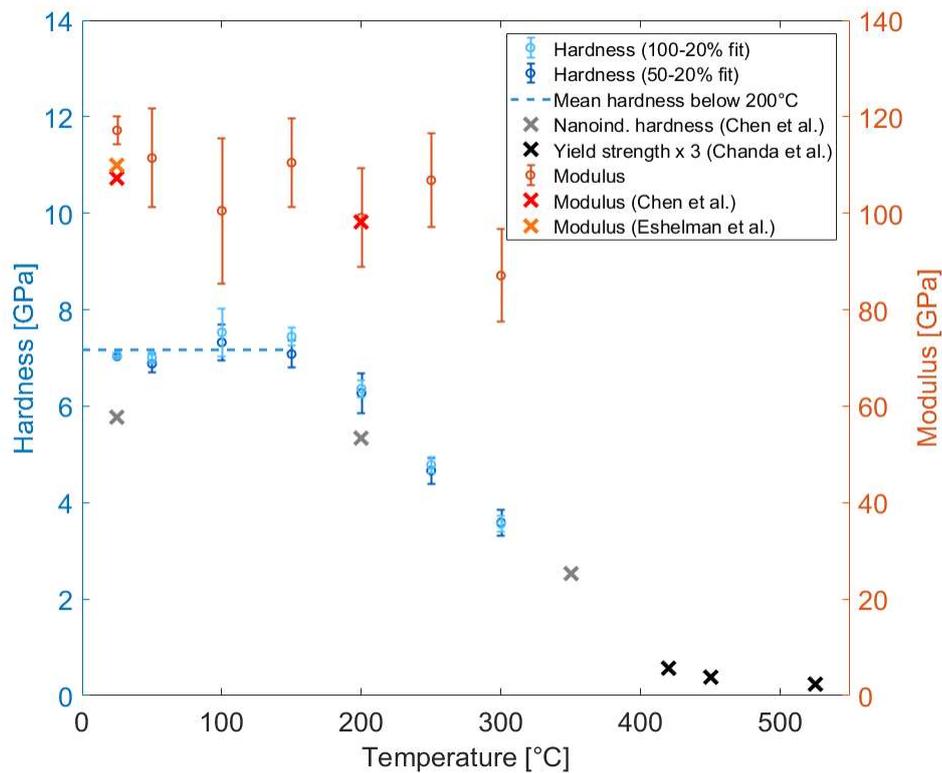

Figure 2: Nanoindentation hardness and reduced modulus of the Al₂Cu θ-phase at different temperatures including their standard deviations. The grey crosses represent the hardness values measured by Chen at al. using nanoindentation [9], while the black crosses are representing the macroscopically measured yield strength by Chanda et al. [1] converted into hardness by multiplication with the factor 3. The red cross represents the modulus values measured by Chen et al. [9], the orange cross represents the modulus measured by Eshelman et al. [5] using the pulse-echo-overlap technique. The dashed blue line corresponds to the mean hardness below 200°C, the temperature where serrations were no longer observed (cf. Figure 1).

In order to reduce the influence of the upper section of the unloading curve on the modulus, the reduced modulus values are only given for the 50-20% fit, Figure 2. The modulus values stay within the standard deviation for the low temperature regime.

For the hardness and reduced modulus values, literature data from Chen et al. [9], who performed nanoindentation tests at three different temperatures, were added to Figure 2 in grey and red for comparison. For the very high temperature hardness, the conversion of macroscopically gained yield stresses in hardness values with the factor three according to e.g. Vandeperre et al. [20] was used and added to the graph (black crosses) [1].

At temperatures below 200°C small serrations were observed during the loading process. These serrations disappear at 200°C and higher temperatures. This is further illustrated in Figure 3 showing the ratio of the actual indenter velocity to the indenter velocity at the previous time step. In this type of plot serrations are displayed by variations in the ratio as present in the low temperature regime. On the other hand, the ratio stays relatively constant in the high temperature regime as indicated by the absence of serrations.

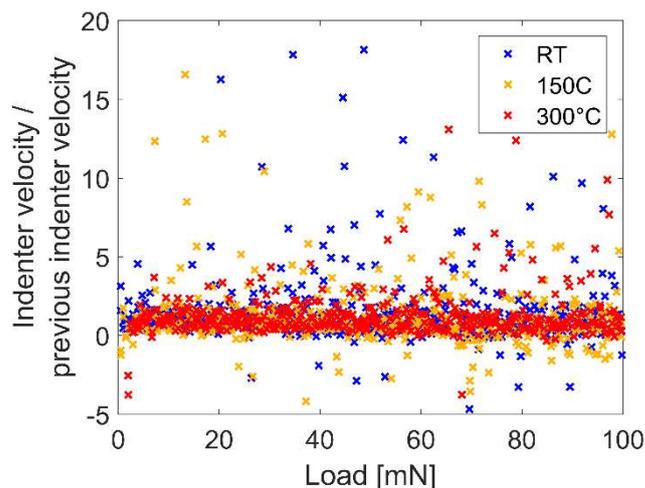

Figure 3: Ratio of the actual indenter velocity to the previous indenter velocity for indents at different temperatures. Significantly higher scatter is seen at low and intermediate temperatures (shown here for the RT and 150˚C data) compared to data obtained at 300˚C, due to serrations in the load-displacement curves.

No slip traces were observed in the vicinity of the indents, neither at the ambient temperature indents, nor at the elevated temperature indents (Figure 4). The surface of indents performed at elevated temperature was further found to be relatively rough (Figure 4 b)), whereas the indent at low temperature shows a relatively flat surface (Figure 4 a)).

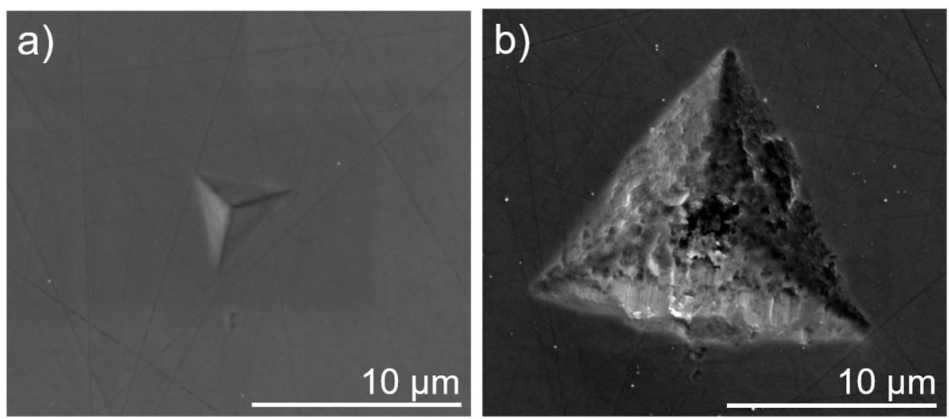

Figure 4: Secondary electron (SE) images of indents in the $Al_2Cu$-θ phase, a) at room temperature, where the indent surface is relatively flat, b) at 300°C with a relatively rough surface.

### 3.2. TEM analysis

Three TEM lamellae were analysed in this work, from indents deformed (i) at room temperature and (ii) at 300°C from the series of indents presented above, and (iii) from an indent placed at room temperature separately, i.e. without any subsequent heating.

The bright field (BF) TEM images of the microstructure under the indents deformed at room temperature and 300°C with subsequent heating are given in Figure 5. The plastic zones of both indents reveal the presence of dislocations as well as areas which are characterised by subgrain-boundaries. The most prominent dislocations for both TEM lamellae are marked by white arrows and visible as long dislocation lines ranging from the upper left corner to the lower right corner.

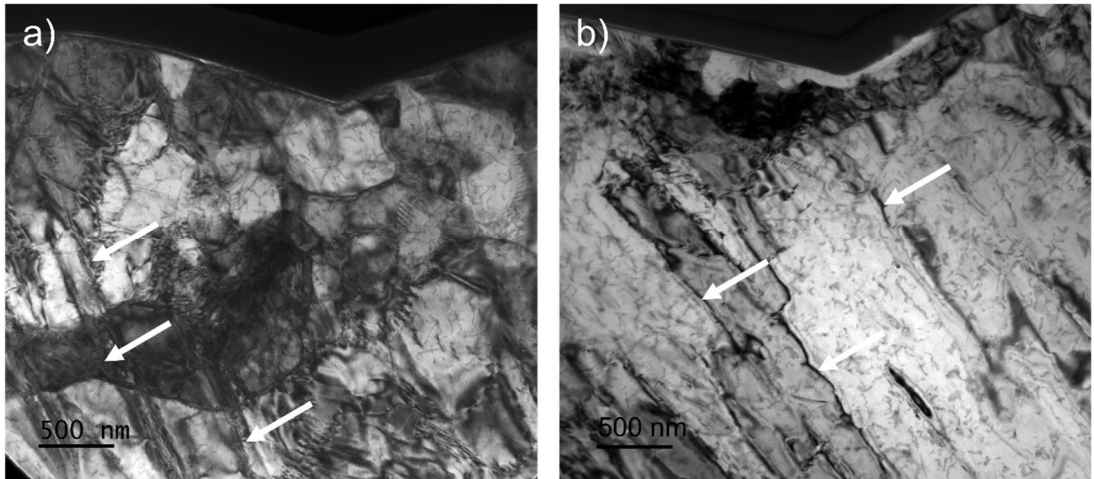

Figure 5: BF images of the microstructure under the indent deformed at a) room temperature and b) 300°C.

The corresponding ambient temperature indent (Figure 5 a)) was further tilted into the [202] zone axis (Figure 6 a)). Here, several dislocations, marked by green, blue and yellow arrows and numbered from 1-3, respectively, are visible. Long segments running nearly vertically across the image are numbered 1, short dislocation segments, running perpendicular to the long dislocation segments are numbered 2 and dislocations lying at an angle of nearly 45° are numbered 3. The corresponding selected area diffraction pattern (SADP) is given in Figure 6 b). When tilting the specimen into the two-beam conditions $(\bar{1}\bar{2}1)$, $(0\bar{2}0)$ and $(\bar{2}\bar{2}2)$, dislocations 1-3 are visible for the diffraction vectors g = $(\bar{1}\bar{2}1)$ and g = $(\bar{2}\bar{2}2)$ but disappear for g = $(0\bar{2}0)$ (Figure 6 c-e)).

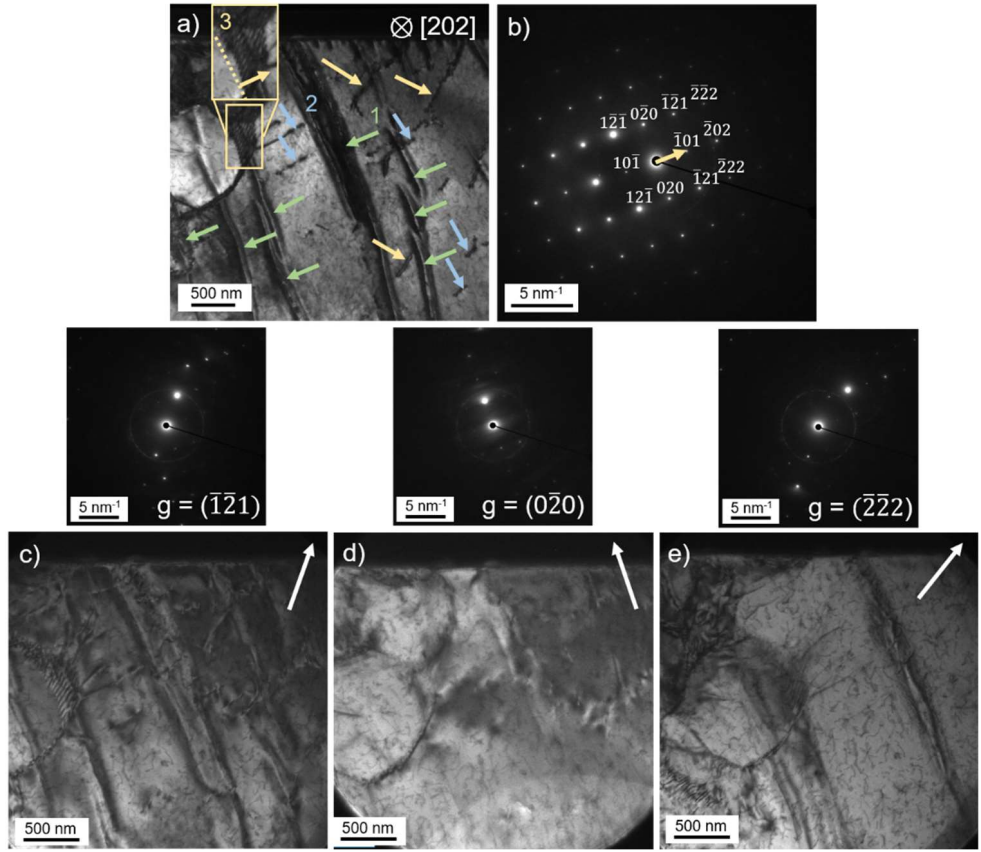

Figure 6: BF-images of the Al$_2$Cu θ-phase deformed at ambient temperature: a) in the [202]-zone axis with dislocations marked in green, blue and yellow including the corresponding diffraction pattern in b). c-e) Three micrographics taken in different two-beam conditions, namely ($\bar{1}2\bar{1}$), ($0\bar{2}0$) and ($\bar{2}2\bar{2}$).

As the dislocation segments of type 1 are very long, we assumed that they lie in a plane that is nearly parallel or only slightly inclined to the thin foil. As the (202) plane is inclined by about 8° from the membrane plane and it is the only low indexed plane within this tilting range, we assume the dislocations of type 1 might lie on the (202) plane. Dislocations of type 2 and 3 probably lie in a plane that is inclined to the membrane as only small segments are present. However, it was not possible to tilt these dislocations into an edge-on orientation. We note that the type 3 dislocations forming the small angel grain boundary in the top left corner of Figure 6, terminate parallel to the dashed line (see Figure 6 a), yellow inset). When we assume that this line is showing the plane trace of these dislocations, the corresponding plane is again of ($\bar{2}02$) type as these planes have a normal vector perpendicular to the plane trace.

In both lamellae heated after and during deformation, subgrains with lower dislocation density were found near the centre of the plastic zone, indicating that the microstructure was affected by the subsequent heating. However, we note that the most prominent dislocations, as marked by the white arrows (Figure 5), are aligned in the same way as the ones seen in the indent placed at room temperature without subsequent heating, presented below.

Figure 7 contains a collage of BF images covering the plastic zone around an indentation placed at room temperature without any subsequent heating.

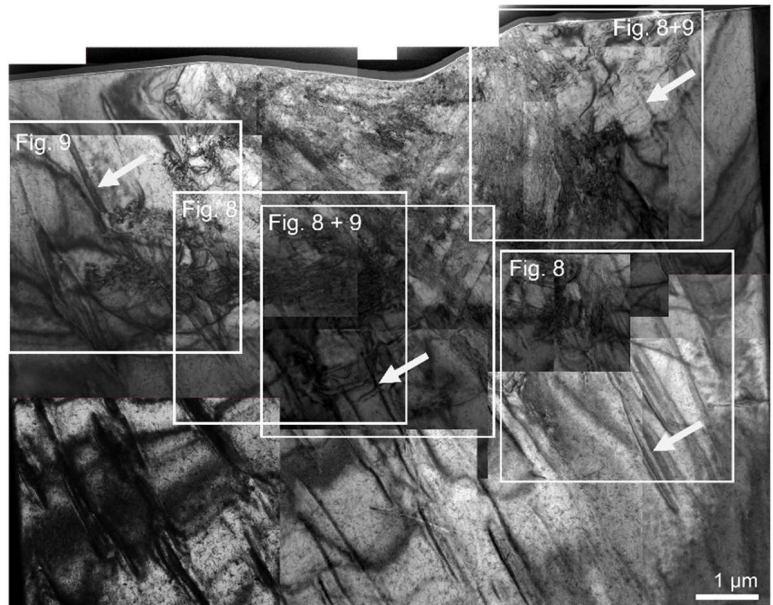

Figure 7: TEM-BF image of the microstructure under the indent deformed at room temperature without subsequent heating. The white arrows indicate the most prominent dislocations in the areas analysed further by two-beam conditions.

The dislocation substructure under this indent was also again analysed using BF images of the [202]-zone axis under different two-beam conditions, namely $(12\bar{1})$, $(0\bar{6}0)$ and $(20\bar{2})$ (Figure 8 a-l)). Again, dislocations are marked by green, blue and yellow arrows and numbered 1, 4, 5 and 6 to refer to these dislocations. Again, long segments running vertically across the image are dislocations of type 1, whereas dislocations of type 4 are approximately 30° inclined and dislocations of type 5 are approximately 45° inclined to the horizontal. Dislocations of type 6 are almost vertically aligned. Due to the same alignment of dislocations of type 1 in the BF images in Figure 6 and Figure 8, the visibility of their long segments as well as their invisibility for $(0\bar{2}0)$ and $(0\bar{6}0)$ two-beam conditions, we assume that these dislocations correspond to the same type.

Dislocations of type 4 are visible for g = $(12\bar{1})$ and g = $(0\bar{6}0)$ and become invisible for g = $(20\bar{2})$, whereas dislocations of type 5 are visible for g = $(12\bar{1})$, invisible for g = $(0\bar{6}0)$ and have a low contrast for g = $(20\bar{2})$. Dislocations of type 6 are visible for g = $(12\bar{1})$ and g = $(20\bar{2})$ but have a low contrast for g = $(0\bar{6}0)$.

Invisible dislocations and dislocations with a low contrast are marked by dashed arrows.

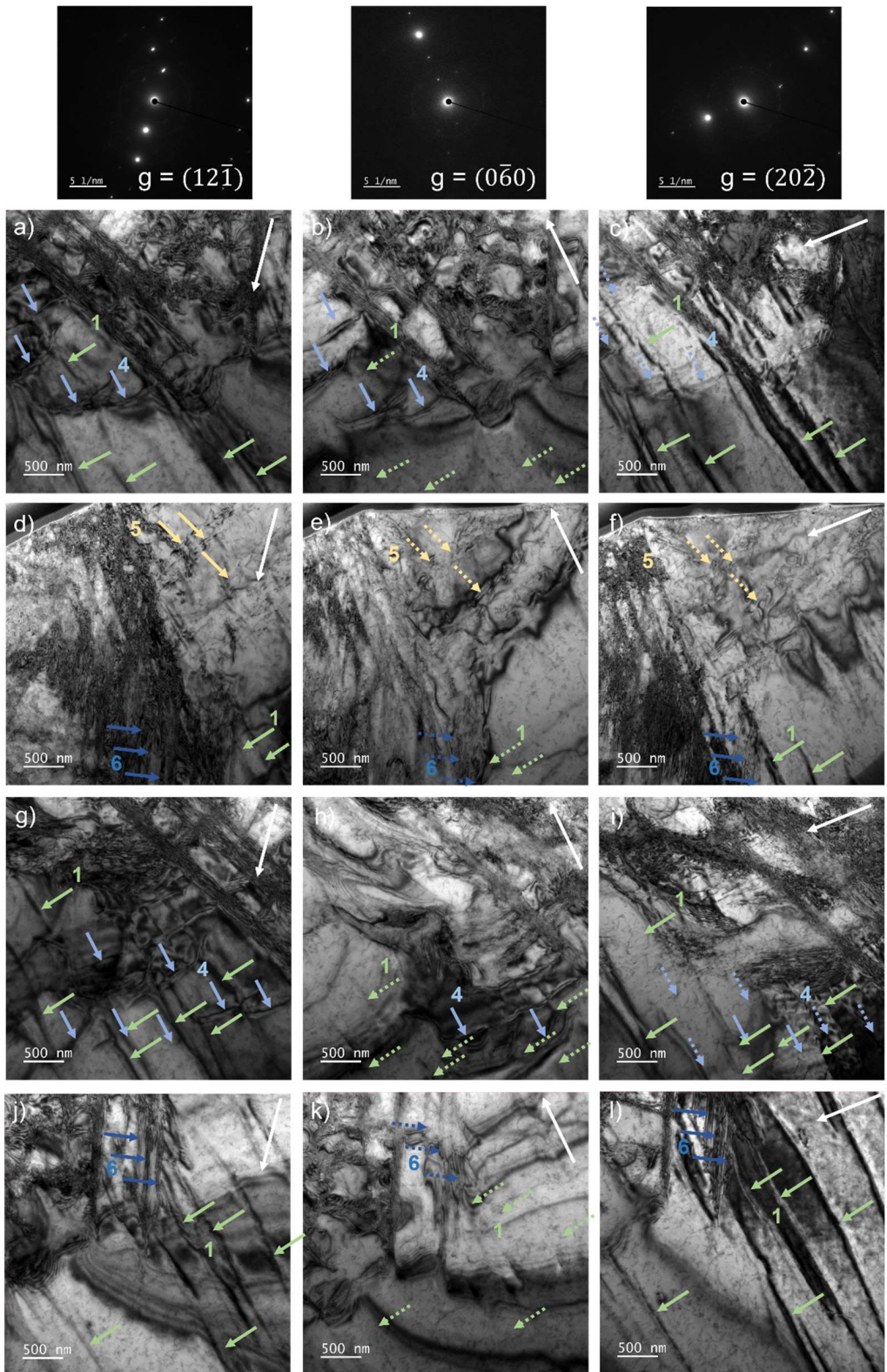

Figure 8: BF-images of the deformed Al₂Cu θ-single phase in the [202]-zone axis. The micrographs in the same row correspond to the same area, whereas the same column corresponds to the same two-beam condition, namely $(12\bar{1})$, $(0\bar{6}0)$ and $(20\bar{2})$. Dislocations of type 1, 4, 5 and 6 are marked by the coloured arrows. If the arrow is dashed, the corresponding dislocation is considered to be invisible.

TEM dark field (DF) and BF-images of the same membrane tilted to the [3$\bar{1}$3]-zone axis are given in Figure 9 a-f). Since the area depicted in Figure 9 a-b) corresponds to the area depicted in Figure 8 a-c), the observed dislocations could be unambiguously identified as type 1 and 4. The same correlation was done for dislocation 5 and 6 in Figure 8 d-f) and Figure 9 c-d).

In Figure 9, dislocations of type 1 are visible for $(\overline{2}02)$, but have a low contrast for the $(2\overline{3}\overline{3})$ two-beam condition, whereas dislocations of type 4 are visible for $(2\overline{3}\overline{3})$ but invisible for $(\overline{2}02)$. Dislocations of type 5 are visible for $(\overline{2}02)$ but have a low contrast for $(2\overline{3}\overline{3})$. Dislocations of type 6 are visible for $(\overline{2}02)$ and for $(2\overline{3}\overline{3})$. Invisible dislocations or dislocations with a low contrast are again marked by dashed arrows.

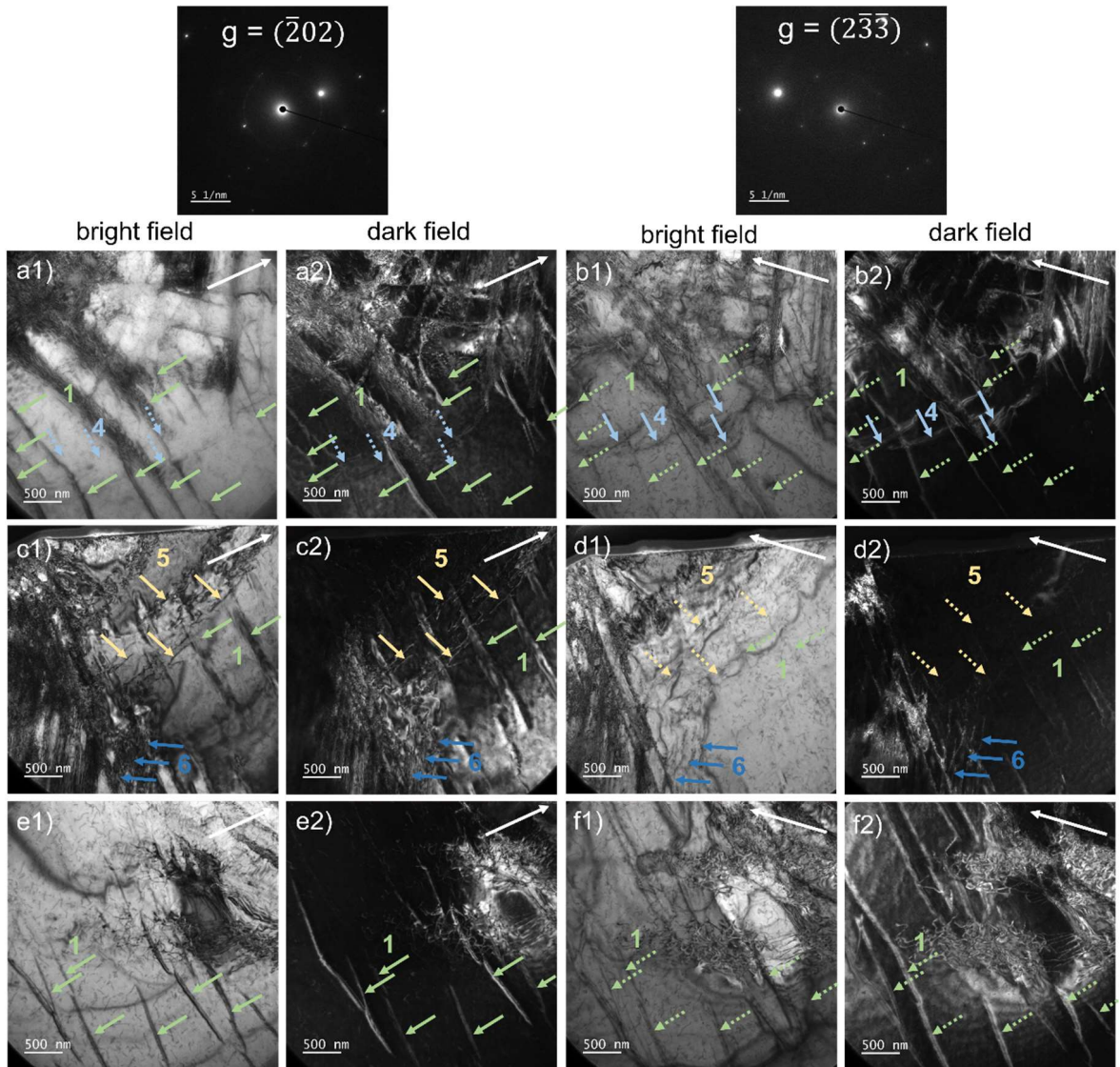

Figure 9: TEM BF- and DF-images of the deformed Al$_2$Cu θ-single phase in the $[3\overline{1}3]$-zone axis including the diffraction pattern. The same row in the micrographs corresponds to the same area, whereas the same column corresponds to the same two-beam condition, namely $(\overline{2}02)$ and $(2\overline{3}\overline{3})$. Dislocations of type 1, 4, 5 and 6 are marked by coloured arrows. If the arrow is dashed, the corresponding dislocation is considered to be invisible.

The dislocations present in the microstructure were analysed using the g·b analysis (b being the Burgers vector) for the eight 2-beam conditions $(\overline{1}\overline{2}1)$, $(0\overline{2}0)$, $(\overline{2}\overline{2}2)$, $(12\overline{1})$, $(0\overline{6}0)$, $(20\overline{2})$, $(\overline{2}02)$ and $(2\overline{3}\overline{3})$. In this analysis, it is assumed that dislocations with different Burgers vectors become visible if the scalar product of g·b ≠ 0 and become invisible if g·b = 0. However, due to nearly no contrast at very small g·b values, dislocations with $g \cdot b \leq 0.33$ cannot be distinguished from $g \cdot b = 0$ and are therefore considered to be invisible [21].

The slip systems as well as potential perfect and dissociated partial dislocations in the $Al_2Cu$ $\theta$-phase are given in Table 1. This data stems from a recent study of the present authors using atomistic simulations [15] revealing that the slip systems highlighted in red are not predicted to occur. They are consequently not considered in the further analysis.

Table 1: Slip systems of the $Al_2Cu$ $\theta$-phase for perfect and partial dislocations. The slip systems marked in red are assumed to be not activated due to a high maximum energy ($E_{max}$) required for shear according to molecular statics simulations. For more information, please read reference [15].

| | Perfect | Partial | |
|---|---|---|---|
| Plane | Direction | Direction | $E_{max}[mJ/m^2]$ |
| {110} | ½ <$\bar{1}$11> | no dissociation | 920 |
| | <001> | ½ <001> + ½ <001> | 795.3 |
| {200} | <001> | ½ <011> + ½ <0$\bar{1}$1> | 1217.6 |
| {002} | <100> | no dissociation | 1296.5 |
| | <110> | ½ <110> + ½ <110> | 1398.2 |
| {211} | ½ <$\bar{1}$11> | $^1/_{10}$ <$\bar{3}$1$\bar{1}$> + $^1/_{20}$ <137> + $^1/_4$ <$\bar{1}$11> | 673.4 |
| {112} | ½ <11$\bar{1}$> | no dissociation | 1065.0 |
| | <$\bar{1}$10> | ½ <$\bar{1}$10> + ½ <$\bar{1}$10> | 1015.5 |
| {310} | <001> | ½ <001> + ½ <001> | 907.6 |
| {022} | <01$\bar{1}$> | $^3/_5$ <01$\bar{1}$> + $^2/_5$ <01$\bar{1}$> | 1113.1 |
| | <100> | $^1/_6$ <211> + $^1/_6$ <2$\bar{1}\bar{1}$> + $^1/_3$ <100> | 789.5 |
| | ½ <11$\bar{1}$> | $^1/_6$ <$\bar{2}$11> + $^1/_6$ <54$\bar{4}$> | 789.5 |

All of these perfect and partial Burgers vectors shown in Table 1 are applied in the present $g \cdot b$ analysis. According to this analysis, potential perfect dislocations of type 1 are indexed to have a [001], [100] or [$\bar{1}$01] Burgers vector component, whereas partial dislocations with [001], [100], [$\bar{1}$01], [$\bar{3}$17] and [31$\bar{7}$] components are possible.

Dislocations of type 2 and 3 are determined to have a [100], [001] or [$\bar{1}$01] Burgers vector component assuming perfect or partial dislocations or a [311], [31$\bar{1}$], [1$\bar{3}$7], [13$\bar{7}$], [$\bar{3}$17], [31$\bar{7}$], [211] or [21$\bar{1}$] component assuming only partial dislocations.

According to the $g \cdot b$ analysis of dislocations of type 4, these dislocations are indexed to either have a [111], [1$\bar{1}$1] or [010] component assuming perfect or partial dislocations or a [544], [5$\bar{4}$4], [454], [4$\bar{5}$4], [211], [121], [1$\bar{2}$1], [131] or [1$\bar{3}$1] component if they were partial dislocations. Dislocations of type 4 occur in conjunction with stacking faults visible in a multibeam bright field condition close to the [202] zone axis (Figure 10) and are therefore thought to be partial dislocations. Due to the additional contrast of the surrounding dislocations obscuring the contrast oscillations in the stacking fault, we were not able to conclusively determine the possible stacking fault type from the two-beam conditions in Figure 8 g) – i).

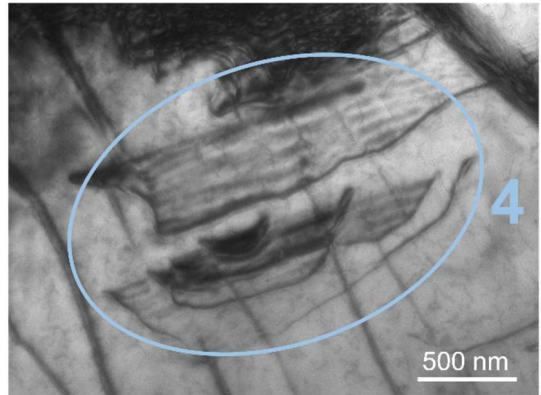

Figure 10: Stacking faults of dislocation type 4 can be seen in multibeam condition close to the [202]-zone axis.

Dislocations of type 5 are determined to have a [100], [001] or [$\bar{1}$01] Burgers vector component when assuming full or partial dislocations or a [$\bar{3}$17] or [31$\bar{7}$] component when assuming partial dislocations.

According to the Burgers vector analysis of dislocation type 6, they are indexed to correspond to [11$\bar{1}$], [100], [001], [011], [$\bar{1}$01], [110] or [1$\bar{1}$0] considering full or partial dislocations or to a [31$\bar{1}$], [$\bar{1}$31], [13$\bar{1}$], [31$\bar{7}$], [1$\bar{3}$7], [13$\bar{7}$], [$\bar{3}$17], [31$\bar{7}$], [211], [21$\bar{1}$], [$\bar{1}$21], [544], [54$\bar{4}$], [$\bar{4}$54] or [45$\bar{4}$] partial, thus no Burgers vector could be excluded.

In order to get a better overview, all of the above-mentioned Burgers vectors, which agree to the current g·b analysis, are marked in green in Table 2.

Table 2: a-e) Potential full and partial Burgers vectors of dislocation 1-6 which are in agreement with the conducted g·b analysis are marked in green. The planes identified as the likely slip planes from trace analysis (see discussion) are also marked in green. The slip systems marked in red are those assumed to possess a high critical resolved shear stress (CRSS), making them less likely to be active, see also Table 1.

### a) Dislocation 1

| Plane | Perfect Direction | Partial Direction |
|---|---|---|
| {110} | ½ <Ī11> | no dissociation |
|  | <001> | ½ <001> + ½ <001> |
| {200} | <001> | ½ <011> + ½ <0Ī1> |
| {002} | <100> | no dissociation |
|  | <110> | ½ <110> + ½ <110> |
| {211} | ½ <Ī11> | ¹/₁₀ <31Ī> + ¹/₂₀ <137> + ¼ <Ī11> |
| {112} | ½ <11Ī> | no dissociation |
|  | <Ī10> | ½ <Ī10> + ½ <Ī10> |
| {310} | <001> | ½ <001> + ½ <001> |
| {022} | <01Ī> | ³/₅ <01Ī> + ²/₅ <01Ī> |
|  | <100> | ¹/₆ <211> + ¹/₆ <2ĪĪ> + ¹/₃ <100> |
|  | ½ <11Ī> | ¹/₆ <2Ī1> + ¹/₆ <54ā> |

### b) Dislocation 2 + 3

| Plane | Perfect Direction | Partial Direction |
|---|---|---|
| {110} | ½ <Ī11> | no dissociation |
|  | <001> | ½ <001> + ½ <001> |
| {200} | <001> | ½ <011> + ½ <0Ī1> |
| {002} | <100> | no dissociation |
|  | <110> | ½ <110> + ½ <110> |
| {211} | ½ <Ī11> | ¹/₁₀ <31Ī> + ¹/₂₀ <137> + ¼ <Ī11> |
| {112} | ½ <11Ī> | no dissociation |
|  | <Ī10> | ½ <Ī10> + ½ <Ī10> |
| {310} | <001> | ½ <001> + ½ <001> |
| {022} | <01Ī> | ³/₅ <01Ī> + ²/₅ <01Ī> |
|  | <100> | ¹/₆ <211> + ¹/₆ <2ĪĪ> + ¹/₃ <100> |
|  | ½ <11Ī> | ¹/₆ <2Ī1> + ¹/₆ <54ā> |

### c) Dislocation 4

| Plane | Perfect Direction | Partial Direction |
|---|---|---|
| {110} | ½ <Ī11> | no dissociation |
|  | <001> | ½ <001> + ½ <001> |
| {200} | <001> | ½ <011> + ½ <0Ī1> |
| {002} | <100> | no dissociation |
|  | <110> | ½ <110> + ½ <110> |
| {211} | ½ <11Ī> | ¹/₁₀ <31Ī> + ¹/₂₀ <137> + ¼ <Ī11> |
| {112} | ½ <11Ī> | no dissociation |
|  | <Ī10> | ½ <Ī10> + ½ <Ī10> |
| {310} | <001> | ½ <001> + ½ <001> |
| {022} | <01Ī> | ³/₅ <01Ī> + ²/₅ <01Ī> |
|  | <100> | ¹/₆ <211> + ¹/₆ <2ĪĪ> + ¹/₃ <100> |
|  | ½ <11Ī> | ¹/₆ <2Ī1> + ¹/₆ <54ā> |

### d) Dislocation 5

| Plane | Perfect Direction | Partial Direction |
|---|---|---|
| {110} | ½ <Ī11> | no dissociation |
|  | <001> | ½ <001> + ½ <001> |
| {200} | <001> | ½ <011> + ½ <0Ī1> |
| {002} | <100> | no dissociation |
|  | <110> | ½ <110> + ½ <110> |
| {211} | ½ <11Ī> | ¹/₁₀ <3Ī1> + ¹/₂₀ <137> + ¼ <Ī11> |
| {112} | ½ <11Ī> | no dissociation |
|  | <Ī10> | ½ <Ī10> + ½ <110> |
| {310} | <001> | ½ <001> + ½ <001> |
| {022} | <01Ī> | ³/₅ <01Ī> + ²/₅ <01Ī> |
|  | <100> | ¹/₆ <211> + ¹/₆ <2ĪĪ> + ¹/₃ <100> |
|  | ½ <11Ī> | ¹/₆ <2Ī1> + ¹/₆ <54ā> |

### e) Dislocation 6

| Plane | Perfect Direction | Partial Direction |
|---|---|---|
| {110} | ½ <Ī11> | no dissociation |
|  | <001> | ½ <001> + ½ <001> |
| {200} | <001> | ½ <011> + ½ <0Ī1> |
| {002} | <100> | no dissociation |
|  | <110> | ½ <110> + ½ <110> |
| {211} | ½ <Ī11> | ¹/₁₀ <3Ī1> + ¹/₂₀ <137> + ¼ <Ī11> |
| {112} | ½ <11Ī> | no dissociation |
|  | <Ī10> | ½ <Ī10> + ½ <Ī10> |
| {310} | <001> | ½ <001> + ½ <001> |
| {022} | <01Ī> | ³/₅ <01Ī> + ²/₅ <01Ī> |
|  | <100> | ¹/₆ <211> + ¹/₆ <2ĪĪ> + ¹/₃ <100> |
|  | ½ <11Ī> | ¹/₆ <2Ī1> + ¹/₆ <54ā> |

## 4. Discussion

### 4.1. Hardness and indentation modulus up to 300 °C

The room temperature hardness obtained in the present study amounts to 7.05 ± 0.06 GPa using the 100-20% fit of the unloading curve data. This agrees well to the value measured by Xiao et al. [8], who reported hardness values in the range between 5 GPa and 7 GPa and Chen et al. [22], who reported an average value of 5.77 ± 0.91 GPa. The hardness values further stayed relatively constant up to a temperature of 150°C. Above 150°C, the hardness values decrease and correspond to $6.36 \pm 0.17$ GPa at 200°C and $3.56 \pm 0.16$ GPa at 300°C using the 100-20% fit of the unloading curve data. These values are slightly higher than those reported by Chen et al. [9] for 200°C (5.33 GPa). When considering that Chen et al. [9] performed indentation tests at a lower loading rate of 1 mN/s and longer dwell time at maximum load of 10 s (compared to the present tests using 10 mN/s and 5 s) and that they indented in a different crystallographic orientation of the $Al_2Cu$ θ-phase, which was reported to have a anisotropic hardness [8], the slight deviations from our results are within the expected experimental error.

Furthermore, serrated yielding was observed in our study below temperatures of 200°C (Figure 3). The occurrence of serrations during ambient temperature nanoindentation was already reported by Xiao et al. [8] and Chen et al. [9, 22]. Chen et al. [9] also observed the disappearance of serrations for their experiments at elevated temperature, similar to the present study, although their transition temperature lay higher at 350°C.

The concurrent disappearance of serrations and change in the hardness slope in the high temperature regime at 200°C (corresponding to $0.55*T_m$) and above, indicate a thermally activated transition in the mechanism controlling or limiting plasticity. This is in good agreement with the results published by Kirsten et al. [23]. In their study, the Brinell hardness of the $Al_2Cu$ θ-phase was measured up to temperatures above 800 K and a loss in hardness starting from $0.62*T_m$ was reported. Further, Dey at al. [2] and later Chanda et al. [1] both performed macroscopic mechanical tests on the $Al_2Cu$-θ phase at high temperatures and reported a large dependence of the flow stress on the temperature. Based on the obtained activation volumes they [1, 2] proposed lattice-controlled flow up to high homologous temperatures, although Dey et al. [2] reported that in addition diffusion-controlled climb might have occurred. In contrast, Ignat et al. [4] observed cross slip in material deformed at 350°C using TEM analyses.

The now higher number of data points allows us to more precisely determine the transition point at 150°C - 200°C between the two regimes of temperature, where serrations at lower temperatures and a decreasing hardness without serrations but with significant creep deformation in the high temperature regime was observed. With this, the gap between the available macroscopic test results above the brittle-to-ductile transformation temperature and the previous data points at intermediate temperatures has now also been closed in terms of the temperature-dependent hardness of the $Al_2Cu$ θ-phase.

The elastic modulus values at room temperature measured within this study amount to $117.16 \pm 2.92$ GPa and are in very good agreement with the existing literature data. Chen et al. [9] reported an elastic modulus at room temperature of 107.3 GPa, whereas Xiao et al. [8] found moduli between 110 GPa and 130 GPa. Further studies by Eshelman et al. [5], who measured the elastic constants using the pulse-echo-overlap technique, revealed a value of 109.9 GPa for the polycrystalline modulus at 4.2 K. DFT calculations performed by Zhang et al. [10] gave an average value of 120 GPa, both corresponding well to our findings. Furthermore, our elastic modulus values slightly decrease with increasing temperature as the elastic modulus at 200°C amounts to $99.11 \pm 10.26$ GPa, consistent with the results reported by Chen et al. [9], who report an indentation modulus of 98.18 GPa.

## 4.2. TEM discussion

In order to assess the underlying deformation mechanisms, transmission electron microscopy was employed to characterize the deformed microstructure. Here, we focus on the sample deformed at room temperature in order to closely correlate the present results to previous results from micropillar compression in conjunction with atomistic simulations [15]. We then compare the room temperature analysis with our preliminary finding of dislocation structures formed during indentation at 300 °C, which will require further in-depth study due to their potential convolution with other mechanisms that are not necessarily deformation-induced, such as recovery or recrystallization.

Both room temperature indents (Figure 5 a) and Figure 7) were analysed using the membrane's crystal orientation and the g·b criterion (Figure 6, Figure 8 and Figure 9). The main defects observed in both indents are the long, linear dislocation segments named type 1 above. We might assume that these dislocations lie in the (202) plane, as it is inclined by only about 8° to the membrane plane. Plastic deformation on this plane of the $Al_2Cu$ θ-phase was already reported by Kirsten et al. [13] after their elevated temperature tests and was also found as primary slip plane in micropillar compression in a recent study by the present authors [15], among other planes. TEM investigations by Ignat et al. [24] further revealed $[\bar{1}00]$ dislocations on the $(0\bar{1}\bar{1})$ plane after creep tests of the eutectic at 350°C. These were assumed to be of screw type since cross slip has been observed. Furthermore, Zhang et al. [25] reported stacking faults on the {011} plane after conducting corrosion tests. Their TEM membrane was also tilted into the [202] zone axis revealing a similar microstructure. Wang et al. [12] reported slip on the {011} plane in the $Al_2Cu$ θ-phase after conducting room temperature deformation tests on the lamellar $Al$-$Al_2Cu$ eutectic. They proposed that the occurrence of this slip system is due to the slip continuity across the $Al$-$Al_2Cu$ interface after conducting an analysis of the corresponding geometric relationships. However, as the aforementioned and the current studies indicate, there is strong evidence that slip continuity might not be the only reason for the occurrence of slip on this plane.

In order to identify the Burgers vector of the dislocations present after deformation at ambient temperature, a g·b analysis was conducted. The analysis revealed that the most prominent dislocations in the microstructure, named type 1, might either have a [001], [100] or $[\bar{1}01]$ Burgers vector component considering full or partial dislocations, whereas partial dislocations with $[\bar{3}17]$ or $[31\bar{7}]$ as Burgers vector component are also in agreement with the images obtained from the different two-beam conditions. Since it is assumed that dislocations of type 1 lie in the (202) plane, their Burgers vector needs to lie in the plane as well. As a consequence, the Burgers vector might correspond to either $[\bar{1}01]$ or the corresponding partials, which are 3/5 $[\bar{1}01]$ + 2/5 $[\bar{1}01]$.

Dislocations with $<\bar{1}01>$ Burgers vector on the {202} plane have, to the best of our knowledge, not been observed yet, while several authors [12, 13, 15, 24, 25] reported slip on {202}. In a recent publication by the authors [15], slip on the {022} plane was only observed in <100> and ½ $<1\bar{1}1>$ directions after micropillar compression tests. The {022}$<\bar{1}01>$ slip system has a smaller effective interplanar/effective Burgers vector ratio than the other {022} slip systems and was therefore concluded to be more difficult to activate. However, due to the complex stress state under an indent, no direct correlation between the slip systems observed in our recent study [15] and the current study can be made.

The g·b analysis of dislocations of type 2 and 3 revealed a [100], [001] or $[\bar{1}01]$ Burgers vector component for perfect or partial dislocations or a [311], $[31\bar{1}]$, $[1\bar{3}7]$, $[13\bar{7}]$, $[\bar{3}17]$, $[31\bar{7}]$, [211] or $[21\bar{1}]$ component when assuming only partial dislocations. However, even though no stacking faults or closely dissociated dislocations are visible in the TEM images for dislocations 2 and 3 (Figure 6), we cannot conclusively distinguish here between perfect dislocations and partial dislocations with a small dissociation width giving rise to a similar total lattice distortion in the two-beam conditions. Since it was not possible to tilt the dislocations edge-on, no information on the corresponding slip plane could be obtained. However, as described in the results section, all dislocations of type 3 are terminating parallel to the dashed yellow line (see Figure 6 a and b). If it is assumed that this line corresponds to the plane trace of the slip system, the plane normal of the slip plane must be perpendicular to the dashed line. This might indicate that these dislocations are lying on a $\{\bar{2}02\}$ plane.

Dislocations of type 4 either have a [111], [1$\bar{1}$1] or [010] component, predicted to occur both as perfect or partial dislocations, or a [544], [5$\bar{4}$4], [454], [4$\bar{5}$4], [211], [121], [1$\bar{2}$1], [131] or [1$\bar{3}$1] component corresponding to partial dislocations. Since stacking faults of dislocation 4 are visible in multibeam condition close to the [202] zone axis (Figure 10 and Figure 11 a)), it is assumed that these dislocations are indeed partial dislocations. All <544> and <211> partial dislocation components can only occur as part of the {022} slip system in either <100> or ½ <111> direction (Table 1), both corresponding to the observed full dislocation components. Since the arising contrast in two-beam conditions usually corresponds to the full dislocation for stacking faults with a small dissociation width, this might indicate that dislocation dissociation occurred on the {022} plane. However, the <111> and <311> partials can also occur as part of the {$\bar{2}$11} ½ <111> slip system (Table 1), which has the lowest maximum energy, occurred in micropillar compression of the Al$_2$Cu θ-phase and was assumed to be the slip system that is easiest activated in a recent study [15]. Since the stacking fault of dislocation 4 terminates parallel to the blue line in Figure 11 a), green inset, a further in-depth analysis of the plane trace with the TEM membrane allowed to identify the plane. The corresponding unit cell including the green membrane plane, assuming the membrane was lying perfectly perpendicular to the electron beam initially, is further given in Figure 11 b). The intersection of the blue (121) plane with the green TEM membrane is aligned with the intersection of the stacking fault with the membrane. It is therefore assumed that the stacking fault lies in the (121) plane. This is further in good agreement with results by Wang et al. [12], who also observed the dissociation of dislocations on this plane in their TEM experiments.

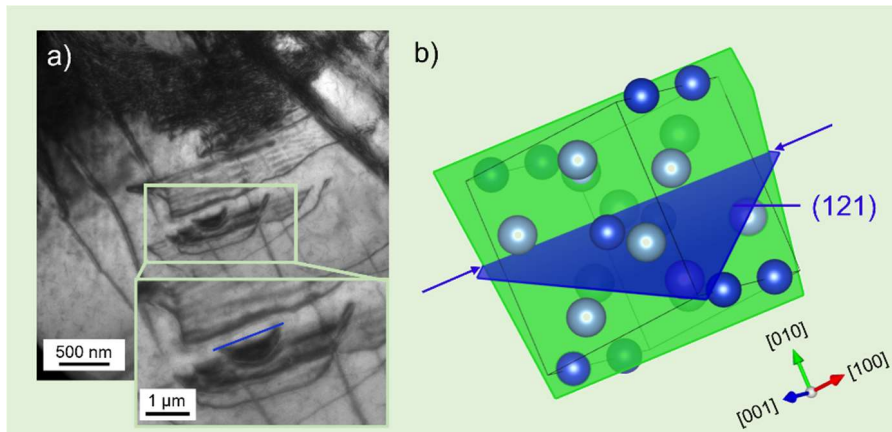

Figure 11: a) TEM BF-image of dislocation 4 including magnified image of the stacking fault, which terminates parallel to the blue line, b) the correspondingly oriented unit cell (generated using VESTA (Visualization for Electron and Structural Analysis, [26])) where the intersection of the (121) plane with the green TEM membrane is highlighted by blue arrows.

Dislocations of type 5 were found to either have a [100], [001] or [$\bar{1}$01] Burgers vector component assuming full dislocations or a partial vector component being a [$\bar{3}$17] or [31$\bar{7}$] partial.

These full Burgers vectors are all part of the {022}<100> slip systems and {022}<01$\bar{1}$> slip systems. Additionally, the {110}<001>, the {200}<001>, the {002}<100> and the {310} <001> slip systems include the observed full Burgers vectors. However, with the available two-beam conditions, no further conclusion can be drawn. Since dislocation 5 is imaged in two zone axes, namely [202] and [3$\bar{1}$3], a closer look onto its plane orientation allowed to identify the underlying slip plane (Figure 12). Here, the unit cell oriented in the corresponding zone axes are depicted together with the (1$\bar{3}$0) slip plane (blue). The plane trace with the green TEM membrane, which is highlighted by blue arrows, corresponds to the plane trace seen in the BF-images in a2) and b2) respectively. Therefore, it is assumed that dislocation 5 corresponds to the (1$\bar{3}$0)<001> slip system or the corresponding partials, which are $\frac{1}{2}$<001> + $\frac{1}{2}$<001>.

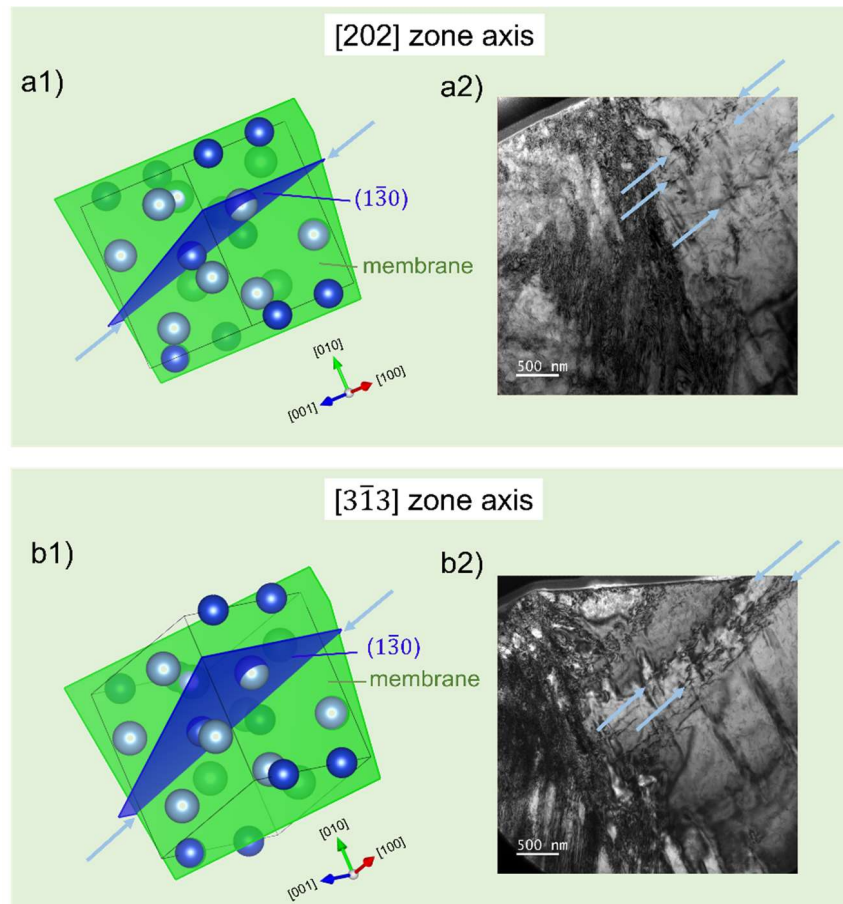

Figure 12: Analysis of the plane trace of dislocation 5 for the two zone axes a) being [202] and b) being [3$\bar{1}$3]. The orientation of the unit cell is given in a1-b1) together with the correspondingly oriented TEM membrane (green) and the (1$\bar{3}$0) plane (blue). a2-b2) represents the corresponding TEM-BF images.

For dislocations of type 6, no closer determination of the Burgers vector was possible applying the g·b analysis with the two-beam conditions reached for this sample (Table 2). However, the plane trace of its slip plane with the TEM membrane in the two zone axes investigated, again being [202] and [3$\bar{1}$3], allowed to identify the ($\bar{1}$12) plane as potential slip plane. The intersection of the green marked TEM membrane with the blue marked ($\bar{1}$12) plane is again highlighted by blue arrows (Figure 13). Since the alignment of these intersections a-b1) is in agreement with the BF-images given in a-b2), it is assumed that dislocations of type 6 lie on the ($\bar{1}$12) plane.

This slip plane was already reported to be activated after macroscopic elevated temperature creep tests of an Al$_2$Cu θ-phase single crystal [3] as well as in a deformed eutectic [4].

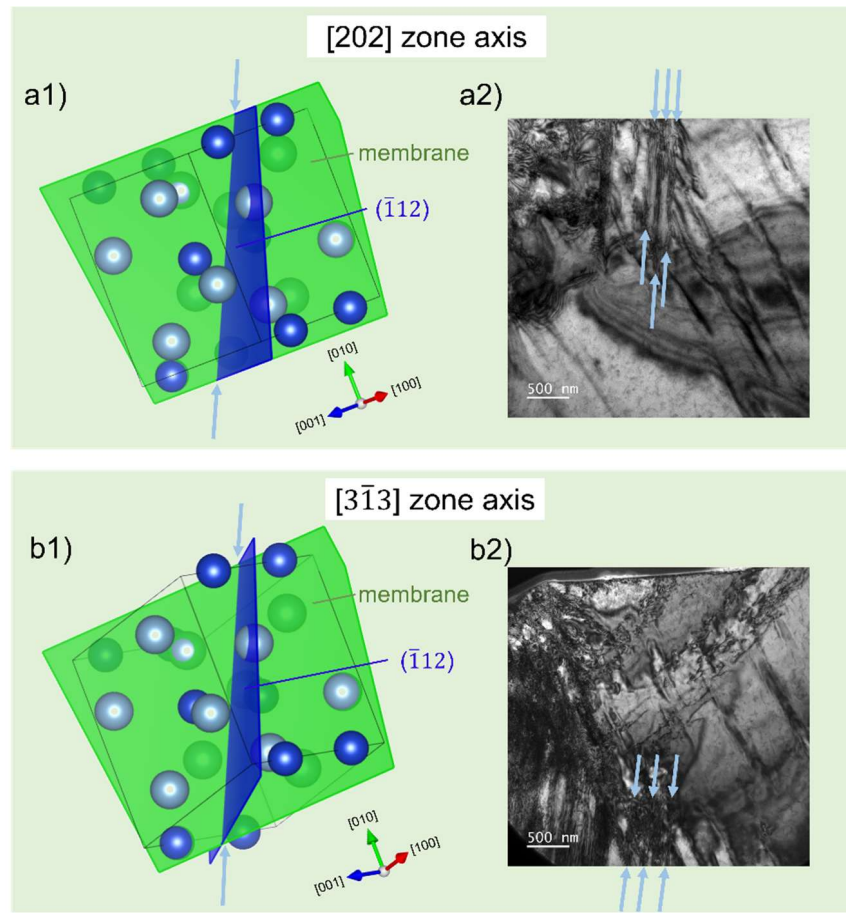

Figure 13: Analysis of plane the trace of dislocation 6 for the two zone axes a) being [202] and b) being [3$\bar{1}$3]. The orientation of the unit cell is given in a1-b1) together with the correspondingly oriented TEM membrane (green) and the ($\bar{1}$12) plane (blue). a2-b2) represents the corresponding TEM-BF images.

The comparison of the characteristic dislocation substructure between ambient and elevated temperature (Figure 5 a-b)), where long segments of dislocations of type 1 are visible close to the [202]-zone axis underneath the indents, indicates that the dislocation-based mechanism of plasticity at both temperatures might operate on the same slip systems. The loss in hardness for temperatures above 200°C may therefore be due to several reasons including a change in dominant slip system with lower activation volume, a change in dislocation core structure or dissociation allowing easier cross-slip of mobile dislocations or a reduced effect of solute atoms limiting flow due to increased diffusion rates.

In addition, the dislocation substructure which is visible in Figure 5 suggests that recovery and recrystallization may also occur. However, it is unclear whether these are dynamic processes or initiated during the long heating and cooling times accompanying the high temperature nanoindentation tests with low drift rates presented here. Consequently, a detailed study of the underlying dislocation mechanisms at high temperature will have to carefully distinguish these phenomena with dedicated experiments using short deformation times and quenching immediately after indentation.

## 5. Conclusions

The plasticity of the Al$_2$Cu θ-phase was investigated from ambient temperature up to 300°C using nanoindentation and TEM investigations. The main findings of this work are:

- At temperatures below 200°C, serrations occur during deformation.
- Thermal softening and a loss in serrations both start at 200°C.
- No slip traces occurred in the vicinity of the indents at ambient and elevated temperature.
- The most prominent dislocations present in the ambient temperature and the elevated temperature indent are assumed to correspond to {202}<$\bar{1}$01> or the corresponding partials, being $^3/_5$ [$\bar{1}$01] + $^2/_5$ [$\bar{1}$01].
- Partial dislocations have been identified and are assumed to lie on the {121} plane.
- Further dislocations corresponding to {1$\bar{3}$0}<001> or {1$\bar{3}$0} ½ <001> + ½ <001> were identified as well as dislocations of unknown Burgers vector on the {$\bar{1}$12} plane.

## Acknowledgement

The authors gratefully acknowledge funding of the priority program "Manipulation of matter controlled by electric and magnetic field: Towards novel synthesis and processing routes of inorganic materials" (SPP 1959/1) by the German Research Foundation (DFG), grant number 319419837. Furthermore, the authors acknowledge funding by the DFG within project A05 of the Collaborative Research Center (SFB) 1394 "Structural and Chemical Atomic Complexity - from defect phase diagrams to material properties", project number 409476157. This project has received further funding from the European Research Council (ERC) under the European Union's Horizon 2020 research and innovation programme (grant agreement No. 852096 FunBlocks).